\documentclass{article}

\usepackage{microtype}
\usepackage{graphicx}
\usepackage{subfigure}
\usepackage{booktabs} 
\usepackage{multirow}
\usepackage{hyperref}
\usepackage{pdflscape}

\usepackage[accepted]{icml2024}

\usepackage{amsmath}
\usepackage{amssymb}
\usepackage{mathtools}
\usepackage{amsthm}
\usepackage{comment}
\usepackage{siunitx}
\usepackage{nicefrac}

\usepackage{enumitem}
\setlist[itemize]{noitemsep,left=7pt,nosep}

\usepackage{caption}
\usepackage[capitalize,noabbrev]{cleveref}

\theoremstyle{plain}

\theoremstyle{definition}

\theoremstyle{remark}

\usepackage[textsize=tiny]{todonotes}

\newcommand\extrafootertext[1]{%
  \bgroup%
  \renewcommand\thefootnote{\fnsymbol{footnote}}%
  \renewcommand\thempfootnote{\fnsymbol{mpfootnote}}%
  \footnotetext[0]{#1}%
  \egroup%
}

\usepackage{bm}
\newcommand{\idx}{\mathbf{i}}
\newcommand{\R}{\mathbb{R}}
\newcommand{\N}{\mathcal{N}}
\newcommand{\vs}{{\bm{s}}}
\newcommand{\vu}{{\bm{u}}}
\newcommand{\vx}{{\bm{x}}}
\newcommand{\vy}{{\bm{y}}}
\newcommand{\vzero}{{\bm{0}}}
\newcommand{\vmu}{{\bm{\mu}}}
\newcommand{\vtheta}{{\bm{\theta}}}
\newcommand{\mI}{{\bm{I}}}
\newcommand{\p}{\mathrm{p}}
\newcommand{\q}{\mathrm{q}}
\newcommand{\Wtke}{W_{2,\mathrm{TKE}}}
\newcommand{\Wr}{W_{2,\mathcal{R}}}

\newcommand{\sigmans}{\sigma_{\mathrm{NS}}}

\icmltitlerunning{Unfolding Time: Generative Modeling for Turbulent Flows in 4D}

\begin{document}

\twocolumn[
\icmltitle{Unfolding Time: Generative Modeling for Turbulent Flows in 4D}



\icmlsetsymbol{equal}{*}

\begin{icmlauthorlist}
\icmlauthor{Abdullah Saydemir}{equal,tum}
\icmlauthor{Marten Lienen}{equal,tum,mdsi}
\icmlauthor{Stephan G\"unnemann}{tum,mdsi}
\end{icmlauthorlist}

\icmlaffiliation{mdsi}{Munich Data Science Institute, Technical University of Munich, Germany}
\icmlaffiliation{tum}{Department of Informatics, Technical University of Munich, Germany}

\icmlcorrespondingauthor{Marten Lienen}{m.lienen@tum.de}

\icmlkeywords{Machine Learning, ICML}

\vskip 0.3in
]



\printAffiliationsAndNotice{\icmlEqualContribution} 

\begin{abstract}
A recent study in turbulent flow simulation demonstrated the potential of generative diffusion models for fast 3D surrogate modeling.
This approach eliminates the need for specifying initial states or performing lengthy simulations, significantly accelerating the process. 
While adept at sampling individual frames from the learned manifold of turbulent flow states, the previous model lacks the capability to generate sequences, hindering analysis of dynamic phenomena.
This work addresses this limitation by introducing a 4D generative diffusion model and a physics-informed guidance technique that enables the generation of realistic sequences of flow states.
Our findings indicate that the proposed method can successfully sample entire subsequences from the turbulent manifold, even though generalizing from individual frames to sequences remains a challenging task.
This advancement opens doors for the application of generative modeling in analyzing the temporal evolution of turbulent flows, providing valuable insights into their complex dynamics.
\end{abstract}

\extrafootertext{Find code and data at \url{https://cs.cit.tum.de/daml/unfolding-time}.}

\section{Introduction}
Computational Fluid Dynamics (CFD) has been a prominent research direction due to vast interest across fields and the challenge of developing new architectures grounded in its broad theoretical framework~\cite{pope2001turbulent}.
Recent advancements in applied AI unlatched fast approximation and analysis tools for fluids, yielding plenty of models that predict the state of particles and the corresponding velocity and pressure fields at a future timestep, autoregressively~\cite{li2020fourier, Chen2020FlowGANAC}.
However, the primary objective of CFD simulations often involves exploring the manifold of all possible flow states rather than determining the exact trajectory of a particular flow~\cite{lienen2024zero}.
\citet{lienen2024zero} demonstrate that a generative model can effectively sample from this manifold directly, eliminating the need for iterative simulations over thousands of steps, saving time and computational resources.

Yet, in many situations turbulent flow simulations are used to investigate the development and decay of vortices, location of separation regions, and reattachment points, i.e.\ properties that are inherently linked to temporal variations.
The model proposed by \citet{lienen2024zero} is capable of providing high-quality snapshots of the flow both visually and according to the physically-motivated metrics they propose.
However, it generates independent snapshots of the simulation instead of sequences, which are necessary to investigate dynamic phenomena.
Furthermore, their model does not explicitly rely on any physical knowledge.

In theory, the physics of the simulation are encoded in the data and, therefore, a perfect generative model would have implicitly learned the Navier-Stokes equation.
In practice, a generative model oblivious to the underlying physics may generate samples that are reasonably close to real data while violating the physics in important ways.
To address such deviations from reality in the samples generated by the model, physical knowledge can be infused into the model in a variety of ways \citep{karniadakis2021physics} such as physically-enhanced loss functions \citep{Oldenburg2022GeometryAP}, integrating the model into a numerical solver \citep{kochkov2021machine}, data augmentation \citep{lu2020extraction} or integrating known equations into the structure of the model itself \citep{lienen2022learning}.

\begin{figure*}
    \centering
    \includegraphics[width=\linewidth]{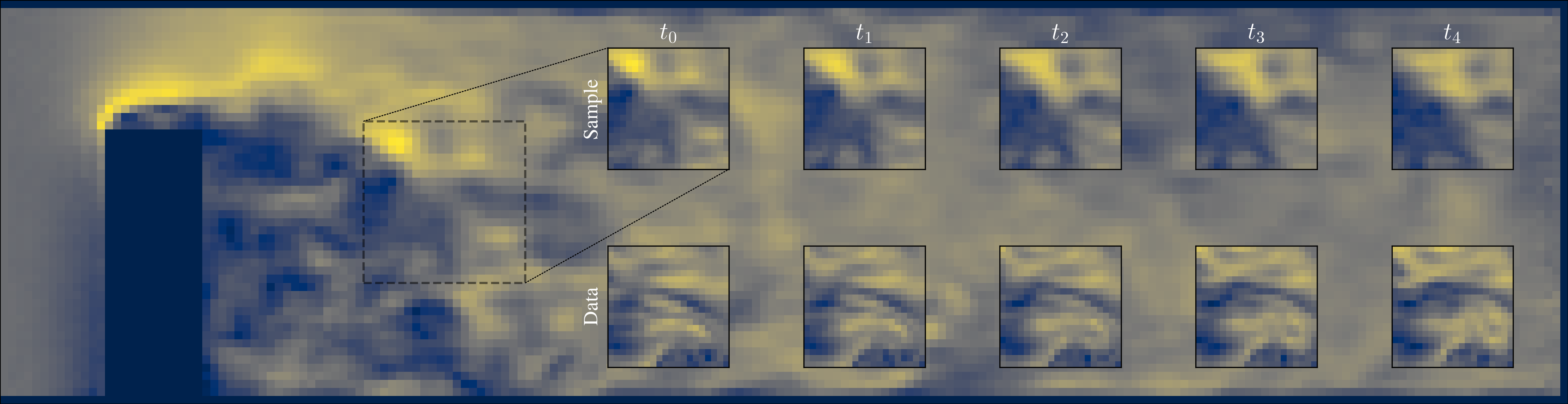}
    \caption{A time-varying 3D turbulent flow simulation generated by our model. Visualization shows the magnitude of the velocity field $\vu$. The insets show that the flow evolves coherently at a similar rate as a ground-truth sample of the same region.}
    \label{fig:zoom-ins}
\end{figure*}

We extend the approach from \citeauthor{lienen2024zero} to generate temporally coherent subsequences of simulations.
In addition to extending the generative approach to simulation to dynamic, time-dependent phenomena, we also propose a physics-informed sampling procedure.
In summary, our contributions are:
\begin{itemize}
    \item We introduce an efficient generative diffusion model for 4D turbulent flow data by combining 3D U-Net with ConvGRU in \cref{sec:method}.
    \item In \cref{sec:guidance}, we derive a physics-informed sampling procedure for diffusion models from the Navier-Stokes equation and classifier guidance.
    \item We show experimentally in \cref{sec:experiments} that simulations generated with our 4D model are of comparable quality to independent snapshots from a 3D model. Our proposed physics-informed guidance improves the sample quality in terms of their turbulent kinetic energy spectra.
\end{itemize}

\section{Background}\label{sec:background}

\subsection{Dataset}\label{sec:dataset}
We use the dataset of 3D flow simulations introduced by \citep{lienen2024zero}.
It consists of 45 simulations in a 0.4x0.1x0.1 cm channel, discretized into a regular grid of 192x48x48 cells.
Each simulation has a differently shaped object placed near the inlet, blocking part of the flow and creating turbulent flow downstream.
The simulations contain \num{0.5}\si{\second} of physical time sampled at regular steps of \num{e-4} seconds for \num{5000} flow states per simulation.
At an inflow velocity of \num{20}\si{\meter\per\second}, the fluid flows 25 channel lengths per simulation and with a low viscosity of $\nu = 10^{-4}$ achieves a Reynolds number of about \num{2e5}, well above the threshold for turbulence.
Yet, the sampling steps are sufficiently finely placed that the average travel distance of the fluid along the channel between two simulation snapshots is just \num{2}\si{\milli\meter}, roughly the diameter of a cell on the simulation grid.
This means that subsequences of the simulation show a coherent evolution of flow states and the dataset is well-suited to the study of dynamical phenomena in the flow.

For each cell $\idx = (i, j, k)$ in the simulation grid and timestep $t$, we have a three-dimensional velocity vector $\vu_{t,\idx} \in \R^3$ and a scalar pressure $p_{t,\idx}$.

\subsection{Generative Turbulence Simulation}\label{sec:gts}
\citet{lienen2024zero} define the task of generative turbulence simulation as sampling from the distribution of all possible flow states 
\begin{equation}
    \p((\vu \parallel p) \mid \mathcal{B}, t \ge t_{\mathrm{turb}}) \label{eq:turbulence-distribution}
\end{equation}
beyond an initial transient phase to turbulence of length $t_{\mathrm{turb}}$ given the boundary conditions $\mathcal{B}$.
Training a generative model $\p_\vtheta$ on this distribution is a viable substitute for numerical simulations to explore the manifold of all possible flow states, because turbulent flows are independent of their initial conditions and ergodic, i.e.\ traverse all possible states in the infinite simulation time limit \citep{galanti2004turbulence}.

\subsection{Generative Diffusion Models}\label{sec:ddpm}
Denoising Diffusion Probabilistic Models (DDPM) are a recently developed class of generative models with great generative capacity and training stability \citep{ddpm,sohl2015deep} that have been extended to diverse domains from molecules \citep{hoogeboom2022equivariant} over adversarial attacks \citep{kollovieh2023assessing} to point processes \citep{ludke2023add}.
DDPMs consist of a forward process $\q(\vx_t \mid \vx_{t-1})$, which transforms a sample $\vx_0 \sim \p$ from the data distribution iteratively into a sample $\vx_T \sim \N(\vzero, \mI)$ of Gaussian noise, and a learned reverse process $\p_{\vtheta}(\vx_{t-1} \mid \vx_t)$, which is trained to reverse the forward process and remove noise.
The training objective is to minimize the KL Divergence between the learnable reverse process $\p_{\vtheta}$ and the forward process $\q$.

\begin{figure*}
    \centering
    \includegraphics{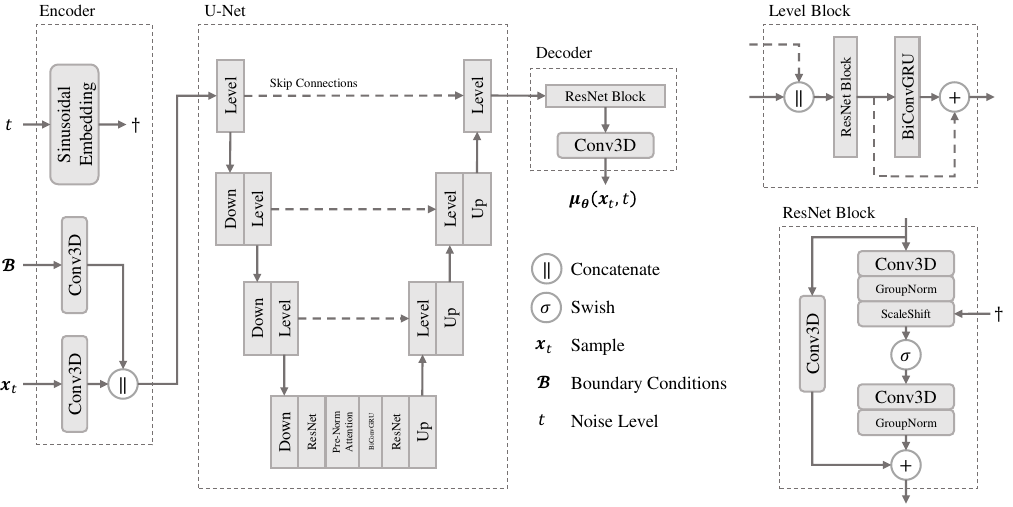}
    \caption{Architecture of our model for the denoised mean prediction $\mu_\vtheta(\vx_t, t)$ in the DDPM framework. The model synchronizes the denoising process along the time dimension at each down- and up-sampling level of the U-Net with bi-directional ConvGRU layers.}\label{fig:architecture}
\end{figure*}

\section{Method}\label{sec:method}
\citet{lienen2024zero} train a generative model to sample turbulent flow snapshots from the distribution in \cref{eq:turbulence-distribution}.
While this allows practitioners to explore the manifold of all possible flow states for some boundary conditions, e.g.\ a car engine design, their approach restricts users to exploring time-independent phenomena in the flow such as surface pressures.
To make their approach viable for the analysis of dynamic phenomena such as mixing times between chemicals, temporally coherent samples are required.

We extend their approach by considering the distribution
\begin{equation}
    \p(\{(\vu_{t'} \parallel p_{t'})\}_{t' = t + i \cdot \Delta t, i = 0\ldots k} \mid \mathcal{B}, t \ge t_{\mathrm{turb}}) \label{eq:coherent}
\end{equation}
of $k$ flow states separated by a time step of $\Delta t$.
In effect, we are training a generative model $\p_\vtheta(\vx)$ for 4D data $\vx_{tijk}$, i.e.\ 3D data varying across time, where $\vx = (\vu \parallel p)$.

To do so, we extend the TurbDiff model from \citet{lienen2024zero}.
TurbDiff is an instance of DDPM that uses a 3D-U-Net \citep{ronneberger2015unet,cicek20163d} to learn the reverse process $\p_\vtheta(\vx_{t-1} \mid \vx_t)$.
At each level of the U-Net, TurbDiff uses 3D convolutions to be parameter efficient.
While such an approach can be directly extended to 4D convolutions in principle \citep{giannopoulos22unet}, it is too costly in terms of runtime and memory.
For the same reason, we also cannot rely on temporal attention as applied by \citet{Ho2022VideoDM}.

Instead, we stay with 3D convolutions and use a bi-directional sequence model along the time dimension.
This can be interpreted as $k$ 3D models communicating along the time dimension to ensure the temporal coherence of the sequence of generated 3D flow states.
The bi-directionality increases the information flow so that state $\vx_{t'}$ and $\vx_{t' + 1}$ can converge jointly towards a sample during the denoising process, instead of $\vx_{t' + 1}$ having to ``follow'' $\vx_{t'}$ as it would be with uni-directional temporal communication.

Between ConvLSTM \citep{shi2015convolutional} and ConvGRU \citep{ballas2015delving}, we choose ConvGRU as our sequence model building block for its lower count of parameters and operations.
Our ConvGRU layer follows the equations in \citet{ballas2015delving}, with all 2D convolutions replaced with 3D convolutions.
We make it bi-directional by applying two independent ConvGRU layers, one in the positive and in the negative time direction, to our sample $\vx_t$ and then combining their outputs with a simple, learned linear layer.

See \cref{fig:architecture} for a visual illustration of our model architecture.

\subsection{Physics-inspired Guidance}\label{sec:guidance}
Diffusion guidance is a mechanism to transform unconditional generative diffusion models into conditional models by guiding the generative process by conditioning the score function \citep{dhariwal2021diffusion}.
If we want to condition our generative model on another random variable $\vy$, we get the updated reverse process \citep{kollovieh2023predict}
\begin{equation}
\begin{aligned}
\p_\vtheta(\vx_{t - 1} \mid \vx_t, \vy) & = \N(\vx_{t - 1} \mid \vmu_\vtheta(\vx_t, t) + \sigma_t^2\vs, \sigma_t^2\mI)\\
\vs & = \nabla_{\vx_t}\log \p(\vy \mid \vx_t)
\end{aligned}\label{eq:guidance}
\end{equation}
where $\vmu_\vtheta(\vx_t, t)$ is the learned denoised mean and $\vs$ is the guiding score.

The Navier-Stokes equation describes the motion of viscous fluids and as such forms the foundation of CFD and numerical solvers like OpenFOAM.
For a given density $\rho$ and viscosity $\nu$, the velocity field $\vu$ and pressure field $p$ have to fulfill
\begin{equation}
    \frac{\partial \vu}{\partial t} = \nu \nabla^2 \vu - \frac{1}{\rho} \nabla p - ( \vu \cdot \nabla) \vu. \label{eq:nse}
\end{equation}
Through guidance, we can use this equation to inject physical knowledge into the generative process as follows.
We define $\vy$ as the residual of \cref{eq:nse},
\begin{equation}
    \vy = -\frac{\partial \vu}{\partial t} + \nu \nabla^2 \vu - \frac{1}{\rho} \nabla p - ( \vu \cdot \nabla) \vu \label{eq:residual}
\end{equation}
and assume that $\vy \sim \N(\vzero, \sigmans^2\mI)$.
An exact solution would have $\vy = \vzero$, but the variance $\sigmans^2\mI$ lets us introduce some slack to account for the unavoidable discretization error inherent to any numerical simulation.

This lets us simplify the guiding score in \cref{eq:guidance} to
\begin{equation}
    \vs = -\frac{1}{2\sigmans^{2}} \nabla_{\vx_t} \|\vy\|^2.
\end{equation}
This form of guidance would not have been possible in the original approach from \citet{lienen2024zero}, because the computation of $\vy$ requires taking gradients with respect to time, which are not available in the generation of independent snapshots.
However, since we are sampling coherent sequences, we can approximate both spatial and temporal derivatives in \cref{eq:residual} with finite differences from the current sample $\vx_t = (\vu_t \parallel p_t)$.

\section{Experiments}\label{sec:experiments}

\begin{table}
    \centering
    \caption{Evaluation results w.r.t. the TKE spectra distance and the regional distribution distance evaluated over sample sets generated from noise with 3 different random seeds.}\label{tab:results}
    \begin{tabular}{cccc}
        \toprule
        Dimension & $\nicefrac{1}{2}\,\sigmans^{-2}$ & $\Wtke$ & $\Wr$ \\
        \midrule
        3 & -- & \num{4.04+-0.08} & \num{1.47+-0.002}  \\
        4 & -- & \num{4.57+-0.08} & \num{1.392+-0.003} \\
        4 & \num{e-12} & \num{4.48+-0.02} & \num{1.326+-0.002} \\
        4 & \num{e-11} & \num{4.47+-0.02} & \num{1.326+-0.002} \\
        4 & \num{e-10} & \num{4.43+-0.02} & \num{1.327+-0.002} \\
        4 & \num{e-9} & \num{4.25+-0.02} & \num{1.332+-0.002} \\
        4 & \num{e-8} & \num{4.25+-0.02} & \num{1.351+-0.002} \\
        
        \bottomrule
    \end{tabular}
\end{table}

We trained two models to evaluate our approach.
The first is a 4D model with the architecture shown in \cref{fig:architecture} trained to generate simulations of length $k = 5$.
To compare against the established approach of generating independent snapshots of the simulation, we trained a second model to generate sequences of length $k = 1$, i.e.\ independent snapshots.
For this model, we replaced all \texttt{BiConvGRU} layers with identity functions.
Both models are trained for 5 epochs and we chose the checkpoint with the lowest $W_{2,\mathrm{TKE}}$ on a validation set for further evaluation.
Our training, validation and test set split follows the setup by \citet{lienen2024zero}.

We evaluate the sample quality of our model with the $\Wtke$ and $\Wr$ metrics proposed by \citet{lienen2024zero}.
The former compares the turbulent kinetic energy (TKE) spectra of the samples and the data, measuring if the kinetic energy of the flow is distributed realistically over the spatial scales.
The latter measures if the samples exhibit the correct velocity, vorticity and pressure distributions in the each region of the simulation domain, e.g.\ low velocities and high pressures in a closed off corner and high velocities in the correct direction at the flow-facing edges of an object.

To compute these metrics, we draw the same number of samples, \num{16} from each model.
Since the metrics compare snapshots invididually, the 4D model would have $16k$ samples with some of them highly correlated compared to the \num{16} independent samples of the 3D model.
To ensure that the metrics can be compared between the models, we select only the middle frame of each sampled sequence for evaluation for the 4D model.

\begin{figure*}
    \centering
    \includegraphics[width=1\linewidth]{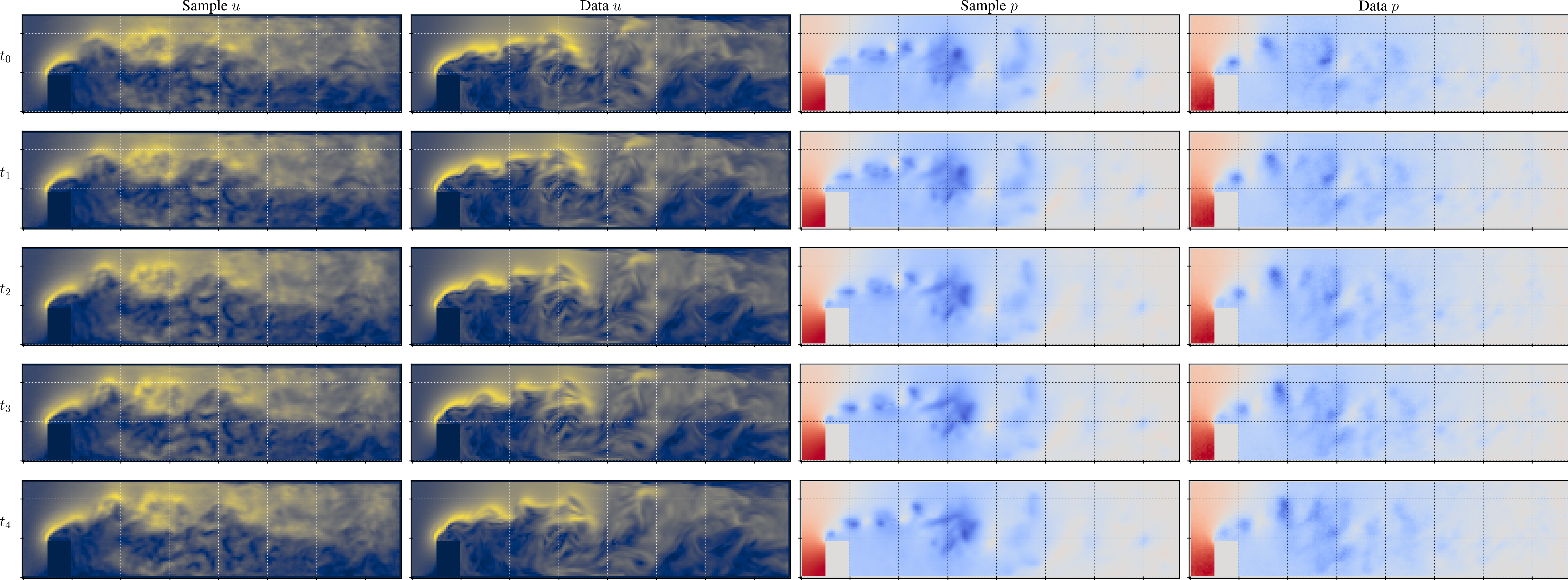}
    \caption{A generated sample for the \texttt{step-high} case compared to a ground-truth example. The dotted grid provides visual support to highlight the dynamics of the flow.}
    \label{fig:sample-step-high}
\end{figure*}

\cref{fig:sample-step-high} shows the velocity and pressure fields from a simulation generated with our model.
See \cref{sec:samples} for additional samples from several test cases.

Our evaluation results in \cref{tab:results} show first that our extended 4D model generates samples of comparable quality to the 3D baseline model.
While the 4D samples are weaker in terms of their TKE spectra, they have a lower regional distribution distance $\Wr$.
This means that the 4D samples are distributing the TKE across the spatial scales less accurately than the 3D samples.
We attribute this to the fact that generating the smallest vortices in a flow seems to be particulary difficult for the model, as we can see in the sample in \cref{fig:sample-step-high}.
However, $\Wr$ being lower in 4D means that the overall distribution of velocities and pressures is more accurate.
Overall, these results show that our 4D model is a viable extension of the 3D baseline model.

We attribute our metrics in the 3D case being overall slightly worse than the ones reported by \citet{lienen2024zero} to the fact that we needed to reduce the model capacity, i.e.\ number of parameters and latent channel dimensions, to ensure that the 4D model could be trained on an NVIDIA A100.

\subsection{Navier-Stokes Guidance}
We also evaluate the effect of our physics-informed guidance (\cref{sec:guidance}) in \cref{tab:results}.
The results show that increasing the strength of the guidance by decreasing the assumed variance $\sigmans^2$ on the residual $\vy$ (\cref{eq:residual}) improves the TKE spectra of the samples while keeping the quality of the overall velocity and pressure distribution constant.
This shows that our guidance is a viable approach to inject physical knowledge into the sampling process of diffusion models.

\section{Limitations \& Future Work}\label{sec:limitations}
The main source of inaccuracy in our physics-informed guidance is the need to approximate the derivatives in \cref{eq:residual} with finite differences.
First, there is the contribution to the inaccuracy of the finite difference approximation from the discretization of the simulation domain.
However, further increasing the resolution of the simulation domain is challenging because the data is already challenging to handle at its current resolution.
Second, the accuracy of the finite difference approximation on a gradient relies on the underlying function being smooth.
However, the simulation data is inherently not smooth because the flow was not fully resolved, which is infeasible in any practical case.

A sophisticated approach to overcome the second source of inaccuracy might be to apply a Large-Eddy-Simulation (LES) technique to the residual in \cref{eq:residual}.
This is the same technique that was used with OpenFOAM to generate the dataset.
In LES, one splits the dynamics into large and small scales and simulates the large scales, while applying a model for the dynamics at sub-grid resolution scale.
A similar approach could enhance the effectiveness of our physics-informed guidance technique further.

\section{Conclusion}
We have extended the generative turbulence simulation approach by \citet{lienen2024zero} to 4D and shown that our model can generate high-quality time-varying simulations of 3D turbulent flows by combining 3D U-Nets with bi-directional ConvGRU layers.
Our Navier-Stokes guidance mechanism lets us inject physical knowledge into the sampling procedure and improves the sample quality.
We believe that our contributions are an important step to make neural networks viable surrogate models for 3D turbulent flows.

\section*{Acknowledgments}
This research was funded by the Bavarian State Ministry for Science and the Arts within the framework of the Geothermal Alliance Bavaria project.


\bibliography{paper}
\bibliographystyle{icml2024}

\newpage
\appendix
\onecolumn

\section{Samples from the 4D model} \label{sec:samples}
Here, we depict samples generated by our model in 4 different scenarios.
Black and white grids are added to aid the visibility of small changes in time.
Exact depictions of the shapes that block the flow are available in the appendix of \citet{lienen2024zero}.

\begin{figure*}[!h]
    \centering
    \includegraphics[width=1\linewidth]{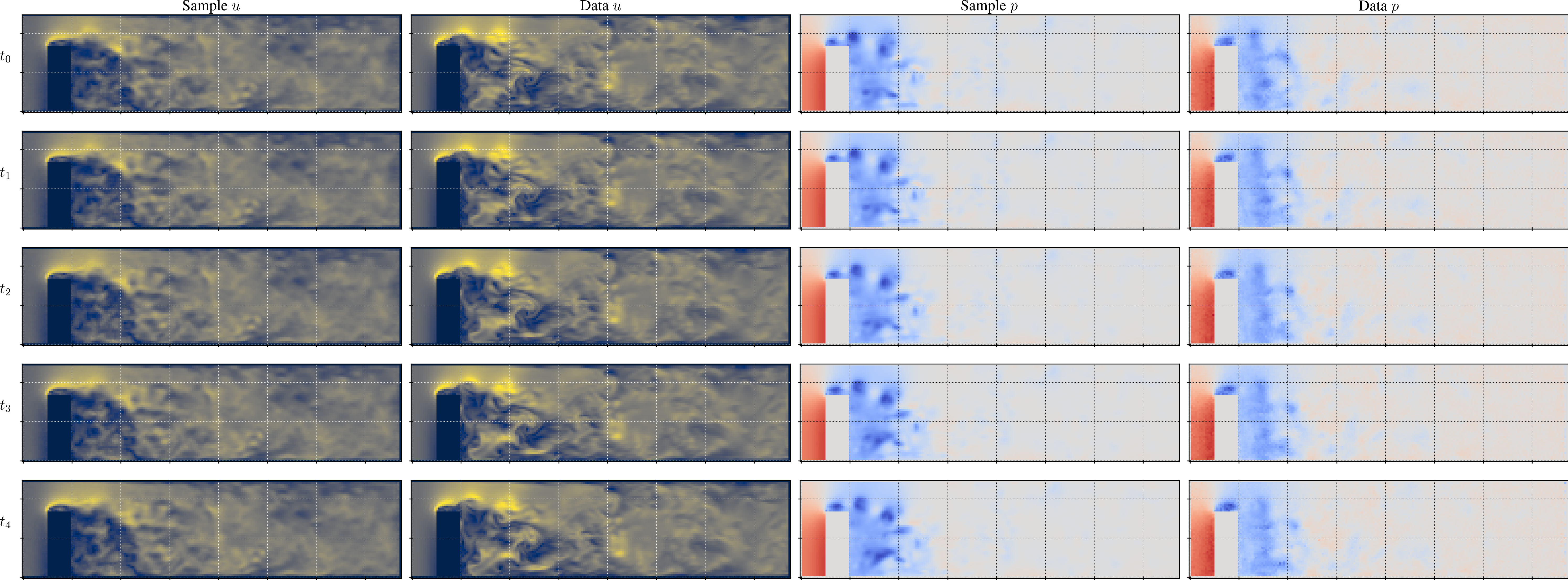}
    \caption{A generated sample from case \texttt{wide-elbow}.~\texttt{wide-elbow} is an ``L" shape rotated 90 degrees clockwise.}
    \label{fig:wide-elbow}
\end{figure*}

\begin{figure*}[!h]
    \centering
    \includegraphics[width=1\linewidth]{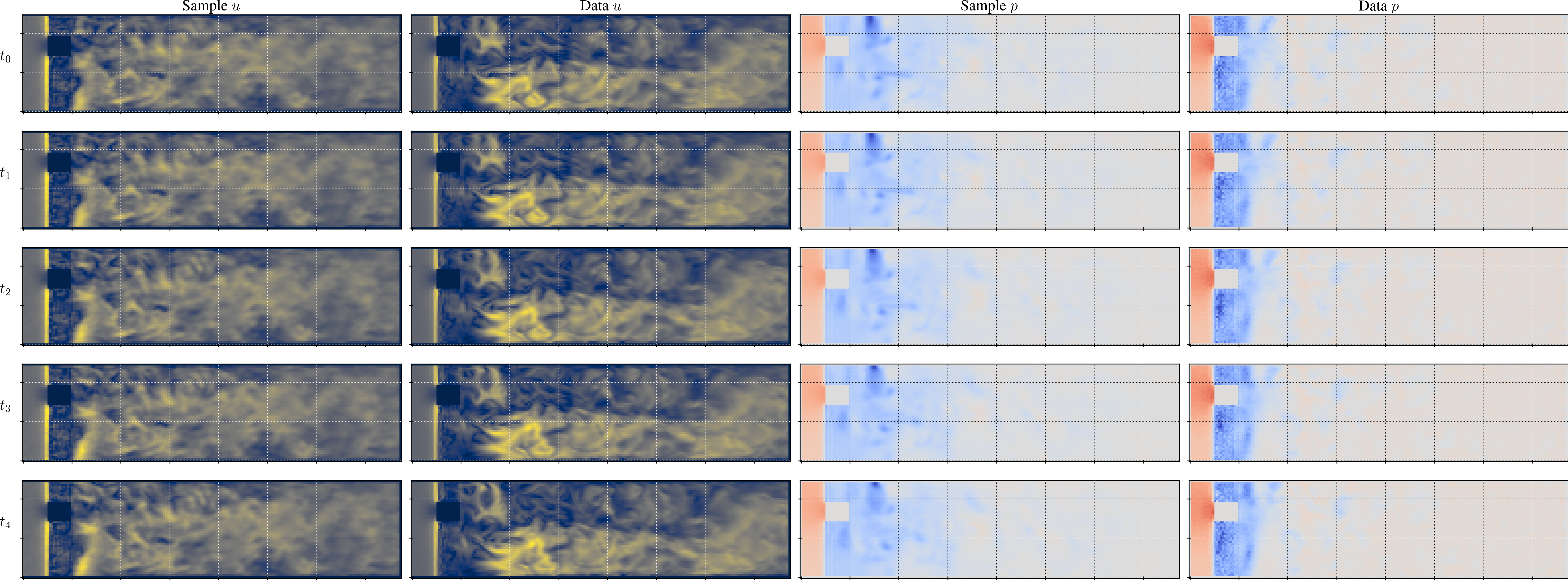}
    \caption{A generated sample from case \texttt{cross-offset}.~\texttt{cross-offset} is a cross but the lines intersect each other on the side rather than in the middle. }
    \label{fig:cross-offset}
\end{figure*}

\begin{figure*}[!h]
    \centering
    \includegraphics[width=1\linewidth]{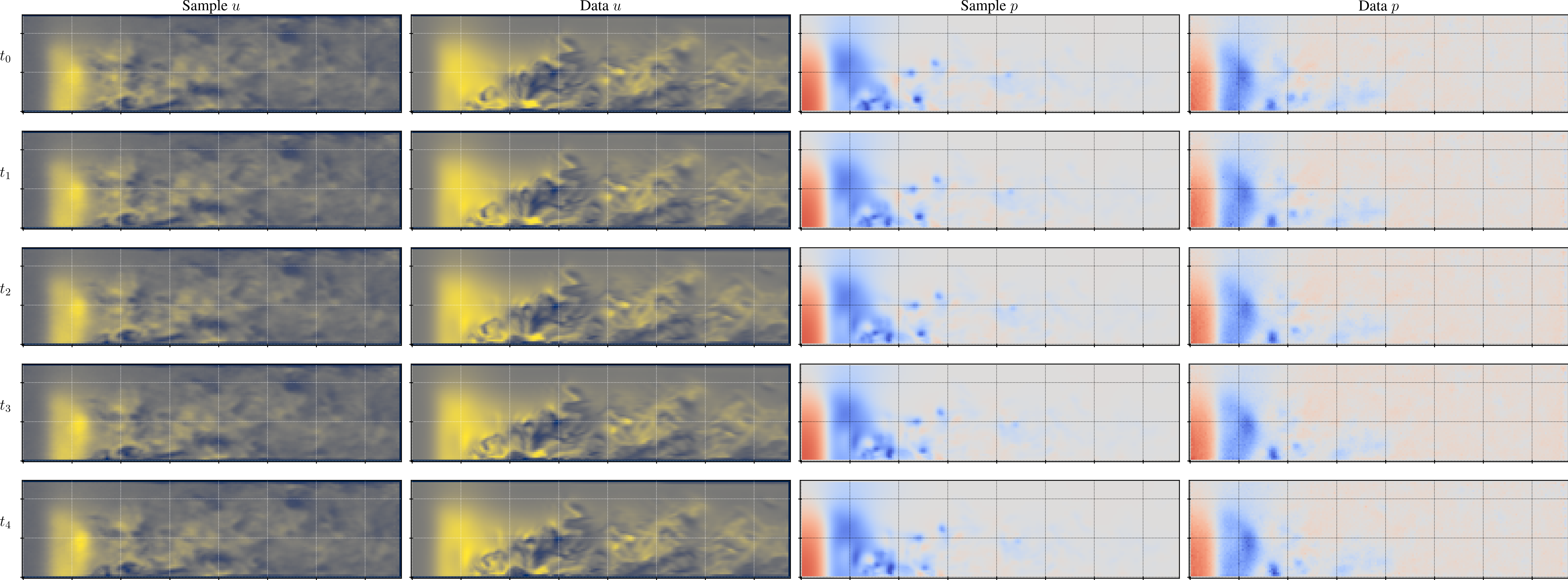}
    \caption{A generated sample from case \texttt{double-pillar}.~\texttt{double-pillar} is two pillars standing side by side. }
    \label{fig:double-pillar}
\end{figure*}

\begin{figure*}[!h]
    \centering
    \includegraphics[width=1\linewidth]{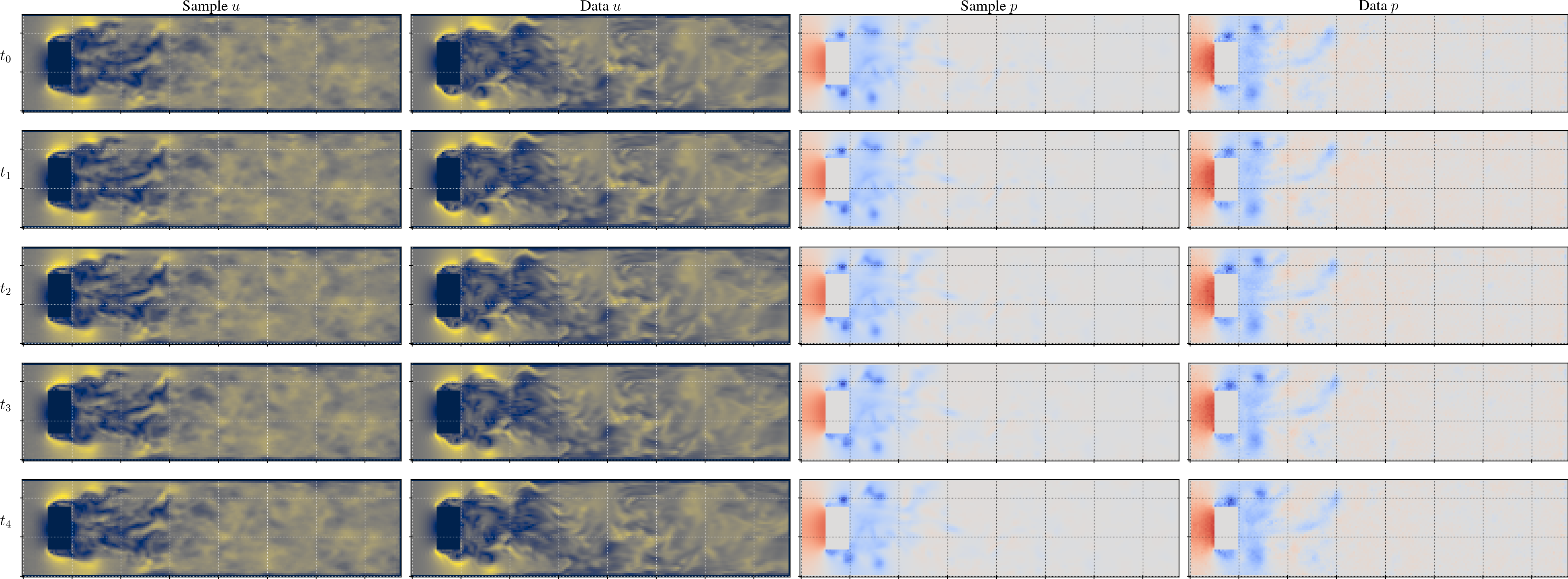}
    \caption{A generated sample from case \texttt{square-large}.~\texttt{square-large} is a large square in the middle of the flow.}
    \label{fig:square-large}
\end{figure*}

\end{document}